\documentclass[twocolumn,showpacs,floatfix]{revtex4}
\usepackage{graphicx}
\usepackage{dcolumn}
\usepackage{amsmath}

\begin{document}
\title{The helix--coil transition on the worm--like chain}
\author{Alex J.~Levine}

\affiliation{Department of Physics, University of Massachusetts,
Amherst. Amherst 01003 MA and The Kavli Institute for Theoretical
  Physics, University of California, Santa Barbara CA 93106}

\date{\today}
\begin{abstract}
I propose a variation of the standard worm--like chain model to
account for internal order parameter (helix/coil) fields on the
polymer chain. This internal order parameter field influences
polymer conformational statistics by locally modifying the
persistence length of the chain.  Using this model, I make predictions for the
bending and stretching response of an alpha-heliacal domain of a protein. In particular, 
I show that alpha-helical protein domains will buckle under applied torque. 
This highly nonlinear elastic behavior  
may be important in the understanding of allosteric control of biochemical pathways.
\end{abstract}

\maketitle

The principal coarse-grained model of a semiflexible polymer is the worm-like chain (WLC).
It describes the single-chain polymer statistics in terms of a quadratic Hamiltonian
that associates an energy cost with chain curvature by introducing a bending modulus
$\kappa$. In thermal equilibrium the effect of this bending energy is to 
enhance the statistical weight of straight polymer conformations on a
length scale $\kappa$ ($k_{\rm B} T = 1$) called
the thermal persistence length\cite{Kratky:49}.  The effect of this bending
energy at longer length scales is trivial and the single-chain statistical mechanics
at those longer scales can once again be described by standard Gaussian statistics
(ignoring for the moment the effect of solvent quality) with an adjusted Kuhn length.

In this letter I propose a coarse-grained polymer model along 
the lines of the WLC designed to describe polymers with 
internal degrees of freedom. In particular, I seek to describe the 
statistical mechanics of alpha-helix forming
domains of proteins\cite{proteins} in an analytically tractable way. By using such a coarse-grained 
model for the (nonlinear) elastic properties of a single protein domain, one can attempt to 
construct a low-dimensional description of protein mechanics (i.e. conformational change) in terms
of a small set of locally interacting subdomains of definite secondary structure. In particular,
a generic consequence of these simple models is that due to the highly nonlinear nature of the effective 
elasticity of an alpha helix, these structures are bistable under an externally applied torque. Their 
mechanical bistability should naturally describe the allosteric control of enzymes\cite{Nelson:04}.

The internal
degrees of freedom are two-state Ising variables that represent the 
presence or absence of hydrogen bonding between the
helix turns of the molecule in its native secondary structure. The 
presence of these {\em internal} state variables -- hydrogen bonding --
couples to the conformational degrees of freedom of the molecule through 
the thermal persistence length. In
the hydrogen--bonded state the polymer is stiffer having a larger bending modulus
since bending the molecule involves the stretching and compression of the hydrogen bonds.

The internal state variable describing the presence/absence of
native secondary structure and consequently local hydrogen bonding
is also controlled by its own energetics in addition to the
aforementioned coupling of these internal degrees of freedom to
the conformational degrees of freedom of the space-curve
representing the polymer chain. The Hamiltonian for these internal
degrees of freedom is a one-dimensional, ferromagnetic
Ising model that in the current context has been referred to as
the ``helix--coil'' (HC) model \cite{Poland:70}. This model has been
previous employed to study a variety of internal state
transformations of biological macromolecules such as the unzipping
of double stranded DNA, and the opening-up of alpha-heliacal
proteins under applied tension \cite{Pincus}. However, this model
has never been combined with the WLC model of the
conformational degrees of freedom of semi-flexible polymers in
order to explore in detail the coupling of these internal degrees
of freedom along the protein backbone with both bending and
stretching degrees of freedom. The new model, the helix--coil
worm--like chain (HCWLC) combines the internal of helix--coil degrees of freedom with
the conformational states of the semiflexible chain.  

The HCWLC Hamiltonian
is a function of discrete set unit tangent vectors $\hat{t}_i, \, i = 1,\ldots
N+1$ and Ising variables $s_i = \pm 1, \, i = 1,\ldots N+1$. In
this coarse--grained description of the polymer, each segment of
the chain has two variables associated with it -- an Ising
variable $s_i$ describing whether that segment of the chain
maintains its native (alpha-helical) secondary structure ($s=+1$)
or not ($s=-1$) and a local tangent vector $\hat{t}$ lying along
the averaged direction of that segment of the chain. See figure
one. Thus, I do not consider more than two internal states of the polymer, but clearly
such a generalization (creating a q-state Potts model
along the 1d chain) is possible. In addition, I do not consider here a
torsional degree of freedom along the chain. In order to reduce the calculation to its 
most fundamental level, I will also analyze the model in two dimensions so that the 
tangent vectors are equivalent to an angular variable. Three dimensional calculations 
follow similarly \cite{Chakrabarti:04}.   Finally, it is important to note that
segments of the chain so described represent more
than a single monomer since these segments must he assigned a
particular, local secondary structure. Based on these
considerations, the Hamiltonian is given by
\begin{eqnarray}
H &=& \frac{\epsilon_{\rm w}}{2} \sum_{i=0}^N \left( 1 - s_i s_{i+1} \right) -
\frac{h}{2} \sum_{i=0}^{N+1} \left( s_i -1 \right) + \nonumber \\
 &  & + \sum_{i=0}^N \kappa \left( s_i \right) \left[ 1 - \cos
\left( \theta_{i+1} - \theta_i \right) \right],
\label{Hamiltonian}
\end{eqnarray}
where the bending modulus of the chain, $\kappa (s)$ takes the form:
\begin{equation}
\label{kap-def}
\kappa(s) = \left\{ \begin{array}{ll}
                        \kappa_\gg & \mbox{if $s = + 1$} \\
                        \kappa_\ll & \mbox{if $s=-1$}
                   \end{array}
            \right.  .
\end{equation}

There are four parameters with the dimensions of energy in
Eqs.~\ref{Hamiltonian},\ref{kap-def}.  The first appearing in the
Hamiltonian is $\epsilon_{\rm w}$ that represents the energy cost
of a domain wall in the internal Ising-like variable. Physically
this energy corresponds to the free energy cost associated with
either starting or ending a sequence of alpha-helical segments of
the polymer and thus reflects a combination of hydrogen--bonding
between consecutive turns of the alpha-helix and the loss of local
entropic configurational free energy due to the adoption of
helical structure. In the standard helix-coil terminology, this
energy is the logarithm of the ``chain cooperativity'' parameter.
The second parameter
in the model is $h$, which appears as a fictitious magnetic field
favoring the spin up ({\em i.e.} alpha-helical secondary
structure) configuration of the chain. In the current
interpretation of the HCWLC model, this term represents effect of
the local, solvent--polymer interactions which provide a
thermodynamic driving force toward its native secondary structure (alpha-helix) 
per unit length. The final two energies enter
in the last term of Eq.~\ref{Hamiltonian}; as seen in
Eq.~\ref{kap-def}, the local bending modulus of the chain takes
one of two values depending on whether that segment of the chain
has its native structure ($s=1$) or not ($s=-1$). Fundamental to
the basic phenomenology discussed below, the bending energy of the
chain is higher in the alpha helical state due to the presence of
hydrogen bonding so that generically $\kappa_\gg \gg \kappa_\ll$.

\begin{figure}
\includegraphics[width=5.3cm]{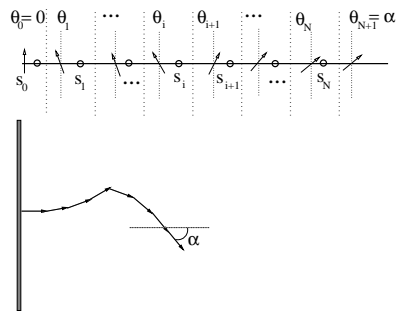}
\caption{The upper figure represents the description of a polymer with 
secondary structure. The open circles represent the internal state 
variable $s_i$ characterizing secondary structure while
vectors are the local tangents to the space curve of the polymer. The 
lower figure shows that space curve (consistent with boundary conditions introduced in 
Eq.~\ref{partition-function}) but does not explicitly show the secondary structure.}
\label{setup}
\end{figure}

One can now compute the response of the chain to applied
torques and forces in equilibrium -- quantities can be directly measured in single-molecule manipulation 
experiments.  I consider the response of molecule to applied
torques which constrain the relative angle between the two end
tangent vectors of the chain to be $\alpha$.  Given that constraint, the partition
function of the polymer is given by
\begin{equation}
\label{partition-function} {\cal Z} (\alpha) = \prod_{i=0}^{N+1}
\sum_{s_i = \pm 1 }\int  d \theta_i
\exp^{-H/T} \delta(\theta_0) \delta(\theta_{N+1} - \alpha),
\end{equation}
where the sum is over all angles and internal states of the chain
while the two delta functions fix
the initial tangent vector to lie along the x axis (removing the
trivial, degree freedom associated with uniform rotations of the
system) and the final tangent vector to make an angle $\alpha$
with the initial one thereby enforcing a particular total bend in
the polymer chain. 

The main qualitative result of this calculation is that, due to the coupling of the
conformation and internal degrees of freedom of the chain, the polymer is bistable under 
applied torque.  Upon increasing the angle $\alpha$, one finds that the necessary torque increases linearly for small angles. Upon reaching a critical 
angle, however, this torque drops dramatically indicating the nucleation of a soft ({\em i.e.} short persistence length) region of the chain. This new instability, which is due to the coupling of the chain persistence length to 
its state of secondary structure, should have a clear experimental signature in single protein manipulation experiments. 

I evaluate the partition function by a transfer matrix technique\cite{Kleinhert:90}.  
Using the identity that relates the exponentiated 
cosine to a sum of modified Bessel functions of integer order,
one can recast the partition sum in terms of a set of variables conjugate to the angles 
$\theta_i, \, i = 1, \ldots N$ (quantized angular momenta $m_i$ of a particle on the 
unit circle)that also form a complete set of states. Using this completeness the partition sum in Eq.~\ref{partition-function} can be written as
\begin{equation}
\label{transfer-form}
{\cal Z} (\alpha) = \sum_{s_0, s_{N+1} = \pm 1} \eta_{s_{N+1}} \langle s_0, \theta_0 \left| 
{\cal T}^{N+1} \right| s_{N+1}, \theta_{N+1} = \alpha \rangle
\end{equation}
where initial ($\langle |$) and final ($| \rangle$) states of the chain are two-component 
objects representing both the secondary structure of the these segments ($s$) and the 
angle of their tangent vector, which has been determined by the delta functions appearing in Eq.~\ref{partition-function}. There is a one-to-one mapping between partition function of a 
chain of $N+1$ segments having its initial and final tangents at angles $0$ and $\alpha$ 
respectively and the quantum transition amplitude of a particle moving on the unit circle 
from angle $0$ to angle $\alpha$ in $N$ time slices. Local secondary 
structure acts like an internal two-state variable 
analogous to the spin state of a spin$-1/2$ particle.  The transfer matrix (or time 
evolution operator) ${\cal T}$ is diagonal in a momentum taking the form of a 
$2 \times 2$ matrix acting on the ``spinor'' space of secondary structure so that:
\begin{eqnarray}
\langle m, s\left| {\cal T} \right| m', s' \rangle &=& \nonumber \\
 \delta_{m,m'} \left( 
\begin{array}{cc}
I_m(\kappa_\gg ) e^{-\kappa_\gg} & I_m(\kappa_\gg ) e^{-\kappa_\gg - \epsilon_{\rm w}} \\
I_m(\kappa_\ll ) e^{-\kappa_\ll - \epsilon_{\rm w}- h } & I_m(\kappa_\ll ) e^{-\kappa_\ll - h}
\end{array}
\right).& &
\label{transfer-matrix}
\end{eqnarray}
The term $\eta_s = \delta_{s,-1}e^{-h}$ in Eq.~ \ref{transfer-form} is a book-keeping device required to 
include the free energy cost the $N+1$ segment of the chain not being in its native state. 
This asymmetry between the two ends of the chain is an artefact of the model that is due 
to the fact that the bending modulus associated with the $n^{\rm th}$ and $(n+1)^{\rm th}$ 
tangent vectors depends on the spin $s_n$ only. Thus the quantum analogue of this problem 
described by the above Hamiltonian is non-Hermitian and consequently asymmetric with 
respect to time reversal.

The transfer matrix defined in Eq.~\ref{transfer-matrix} can now be trivially 
diagonalized in the spinor space of secondary structure by a similarity transformation. 
The eigenvalues of the transfer matrix are given by
\begin{equation}
\label{eigenvalues}
\lambda_{1,2}(m) = \frac{\omega_m (\kappa_\gg)}{2} \left\{ 1 + z_m \pm \sqrt{ 
\left( 1 - z_m \right)^2 + \beta^2 z_m } \right\},
\end{equation}
where $\omega_m(\kappa) = \exp({-\kappa}) I_m(k\kappa)$ is the scalar operator that 
plays the role of the transfer matrix for the all alpha helical, WLC,
while $z_m = \exp({-h}) \omega_m(\kappa_\ll)/ \omega_m(\kappa_\gg)$ 
is the ratio of the fugacity of a random coil (non-native state) 
segment to the alpha-helical fugacity both evaluated at angular momentum $m$. Finally, 
$\beta$ is the exponentiated free-energy cost of introducing a domain wall in the spinor 
variables representing a change in the local secondary structure: $\beta = \exp ( 2 - \epsilon_{\rm w})$, 
where the first term reflects the increase in entropy associated with the appearance 
of one domain wall, and the second term is the energetic cost of such a structure.

The diagonalization of the transfer matrix has been effected by the similarity 
transformation: ${\cal T}^{N+1} \longrightarrow {\cal U}^{-1} \cdot {\cal T}^{N+1} {\cal U}$. 
In the thermodynamic limit $N \longrightarrow \infty$, one may impose cyclic boundary 
conditions and then use the properties of the trace to eliminate the matrix ${\cal U}$ 
with its inverse. Here, it is important to keep $N$ finite to study the effect of the 
degree of polymerization $N$ on the phenomenology of the model. In addition, 
alpha-helical segments of proteins typically contain no more than order ten alpha-helical 
turns so the quantitative application of the model to real protein mechanics 
demands an accurate description of the system at these small degrees of polymerization. 
Using the eigenvalues defined in Eq.~\ref{eigenvalues} we write the partition function as 
\begin{eqnarray}
{\cal Z} (\alpha) &= & \sum_{m=-\infty}^{\infty} e^{i \alpha m}  \left\{ \frac{1+ e^{-h}}{2} \left[ \lambda_1^N(m) + \lambda_2^N(m) \right] + \right. \nonumber \\
&+& \left.  \frac{\lambda_1^N(m) - \lambda_2^N(m)}{2 \sqrt{(1-z_m)^2 + \beta^2 z_m }} \cdot \right.\nonumber \\
& \cdot & \left. \left(1 + \beta z_m - z_m - e^{-h} \left[1- z_m - \beta \right] \right) \right\}.
\label{final-partition}
\end{eqnarray}
One can gain insight into this expression in the limit of high chain cooperativity 
where $\beta \ll 1$, $\lambda_2(m) \ll 1$ by expanding in powers of $\beta$.
By collecting terms in this way
${\cal Z} = {\cal Z}^{(0)} + \beta {\cal Z}^{(1)} + \cdots$ we reorganize the 
partition sum into a term counting all conformational states of the all 
alpha-helix and all random coil WLC (${\cal Z}^{(0)}$) plus all 
states of the polymer with one domain wall (${\cal Z}^{(1)}$) {\em etc.}  
    
The second term of Eq.~\ref{final-partition}
is due to the nonperiodic boundary conditions and is the sole source of 
such  odd-order $\beta$ terms. Viewing the appearance 
of random coil segments of the chain as a nucleation problem, one may refer to $\beta^1$ 
terms as representing a form of heterogeneous nucleation; in this limit the domain 
wall energy is so high that ``melting'' the alpha-helical structure from the chain 
ends is energetically favorable. For longer chain lengths and/or for smaller domain 
wall energies, melting internal segments becomes favorable due to increased translational 
entropy associated with the placement of random--coil segment compared to the 
energetic cost of a second domain wall. This transition to internal, or homogeneous 
nucleation of random--coil segments should occur at chain lengths where 
$\log N \sim \epsilon_{\rm w}$. For alpha-helical domains of {\em e.g.} six helical 
turns, this transition between the two forms of melting occurs at domain wall 
energies around $1.5 k_{\rm B} T$.

\begin{figure}
\includegraphics[width=5.5cm]{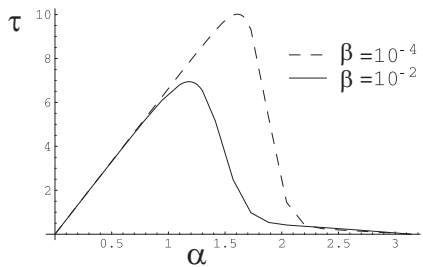}
\caption{The upper (dashed) curve shows the torque $\tau$ (measured in units of $k_{\rm B} T$ vs.\ angle $\alpha$ curve for the model with $\beta = 10^{-4}$. The lower (solid) curve is the equivalent plot for $\beta= 10^{-2}$.  For both curves, $N=15$, $\kappa_\gg = 100$, $\kappa_\ll = 1$, $h=3$. }
\label{torque-fig}
\end{figure}

Based on the partition function given in Eq.~\ref{final-partition}, we can 
compute all statistical properties of the chain for which the directions of 
the ends are constrained. We two most interesting quantities are: (1) the 
mean torque required to fix a given angular bend of the chain $\tau(\alpha)$ 
and (2) the fraction of the chain in the non-native (random coil) state as 
function of the same angle $M(\alpha)$. These quantities are given respectively by
\begin{eqnarray}
\label{torque}
\tau(\alpha) &=& \frac{\partial  \ln {\cal Z}}{\partial \alpha} \\
M(\alpha ) &=& - \frac{1}{N} \frac{\partial \ln {\cal Z}}{\partial h}.
\label{magnetization}
\end{eqnarray}
Both of these quantities can be expressed as sums over one remaining 
angular momentum by taking the appropriate derivatives of Eq.~\ref{final-partition}. 
These infinite sums are exact results. The torque angle curve (approximated by a partial sum) is shown in figure \ref{torque-fig}. Keeping the first
fifteen terms gives this curve with an accuracy of one part in $10^6$. 

The principal qualitative feature of these torque angle curves is that for small angles the $\alpha$-helical chain bends as a simple, semiflexible object; at a critical angle, however, the free energy of the chain is minimized by the creation of a random-coil segment. The free energy increase associated with the introduction of the thermodynamically unfavored random-coil segment is offset by the decrease in curvature 
energy due to the fact that the chain can now bend at the more flexible link thereby created.  
It may be noted that for the physically relevant
quantities used in figure \ref{torque-fig}, the critical torque is on the scale 
$5 - \, 10 k_{\rm B}T \approx 20 - \, 40 \mbox{pN} \cdot \mbox{nm}$; such torques are clearly 
achievable through {\em e.g.} molecular motors acting over length scales of merely 
a few nanometers.  One can observe the creation of random--coil segments by computing 
the magnetization, or fraction of the chain with native secondary structure.  At the buckling
angle (the maxima of the 
curves in figure \ref{torque-fig}), $M \sim {\cal O}(1/N)$ demonstrating the nucleation a single weak link. 
The complete set of curves will be shown later\cite{Chakrabarti:04}.

Another testable experimental prediction involves the force vs.\ 
extension curves of such a molecule. Such calculations have been performed for the WLC model\cite{Marko:95,Lamura:01}. While the torque angle curves can be simply computed in an exact manner due to 
the fact that the angle variables form a natural representation of the semiflexible chain, the force extension behavior of the model is somewhat less transparent. These calculations can, however, be preformed perturbatively 
in the strength of the applied force. The simple structure of the 
perturbation series (shown below) allows the efficient computation of terms to arbitrary order in the applied force. To introduce applied forces, we include a potential in the Hamiltonian $H$ from Eq.~\ref{Hamiltonian}so that
\begin{equation}
\label{modified-Hamilitonian}
H \longrightarrow H - f \sum_{i=0}^{N+1} \gamma(s_i) \cos \theta_i
\end{equation}
where $\gamma(s)$ is the length of a segment, which depends on the local secondary 
structure: $\gamma(+1)  < \gamma(-1)$ and $f$ is the applied force 
stretching the end of the chain in the $\hat{x}$ direction. The exact evaluation of the 
partition sum is now quite complex since the complete set of states needed to insert 
between time-evolution operators is now the wave function of the quantum pendulum (
Matheau functions). The extensional response of the chain to an applied force,however, can be computed perturbatively in powers of $f$:
\begin{eqnarray}
{\cal Z}(\alpha, f)  &= & {\cal Z}(\alpha, f=0) + f \sum_{i=0}^{N+1} \langle \gamma(s_i) \cos \theta_i \rangle_0 + \\
\nonumber
& & + 
\frac{1}{2} f^2 \sum_{i=0,j=0}^{N+1} \langle \gamma(s_i) \cos \theta_i   \gamma(s_j) \cos \theta_j \rangle_0
+ \cdots
\label{perturbation-series}
\end{eqnarray}
where the thermal averages $\langle \cdot \rangle_0$ are taken with respect to the Hamiltonian with $f=0$ \cite{Chakrabarti:04}. The computation of these averages is straightforward using the transfer matrix method.

HCWLC theory is the simplest extension of the WLC admitting a coupling of the chain conformational degrees of freedom to internal state variables. It replaces one persistence length of the WLC with four energy scales that must be determined via experiment, however, it generically creates bistable configurations of the chain under applied torque. The bistability of the individual subdomains of proteins provides a framework for understanding protein conformational change.

The author thanks Fyl Pincus, G.F. Fredrickson, R. Philips, M. Muthumkumar, and K. Dill for 
numerous discussions.  This work was supported in part by the National Science Foundation under grant number DMR98-70785.

\end{document}